\documentclass[usenatbib]{mn2e}
\usepackage{aasmacros,amsmath,amssymb,graphicx}

\def\hmsol{\ensuremath{\,h^{-1}\,\mbox{M}_\odot}}

\def\hmpc{\ensuremath{\,h^{-1}\,\mbox{Mpc}}}

\def\hkpc{\ensuremath{\,h^{-1}\,\mbox{kpc}}}

\def\ms{M_{\rm sat}}
\def\mh{M_{\rm host}}
\def\mvir{M_{\rm vir}}
\def\rvir{R_{\rm vir}}
\def\vvir{V_{\rm vir}}

\def\vt{V_{\rm \theta}}
\def\vr{V_{\rm r}}
\def\vtot{V_{\rm tot}}
\def\rperi{r_{\rm peri}}
\def\rapo{r_{\rm apo}}

\begin{document}
\title{On the Orbits of Infalling Satellite Halos}
\author[Wetzel]{Andrew R. Wetzel$^1$\\
$^1$Department of Astronomy, University of California, Berkeley, CA 94720, USA}

\date{September 2010}

\pagerange{\pageref{firstpage}--\pageref{lastpage}} \pubyear{2010}

\maketitle

\label{firstpage}

\begin{abstract}
The orbital properties of infalling satellite halos set the initial conditions which control the subsequent evolution of subhalos and the galaxies that they host, with implications for mass stripping, star formation quenching, and merging.
Using a high-resolution, cosmological $N$-body simulation, I examine the orbital parameters of satellite halos as they merge with larger host halos, focusing primarily on orbital circularity and pericenter.
I explore in detail how these orbital parameters depend on mass and redshift.
Satellite orbits become more radial and plunge deeper into their host halo at higher host halo mass, but they do not significantly depend on satellite halo mass.
Additionally, satellite orbits become more radial and plunge deeper into their host halos at higher redshift.
I also examine satellite velocities, finding that most satellites infall with less specific angular momentum than the host halo virial value, but that satellites are `hotter' than the host virial velocity.
I discuss the implications of these results to the processes of galaxy formation and evolution, and I provide fitting formulas to the mass and redshift dependence of satellite orbital circularity and pericenter.
\end{abstract}


\section{Introduction}

In hierarchical structure formation, dark matter halos grow via accretion of both diffuse matter and virialized satellite halos.
Beyond driving just halo mass growth, the nature of satellite accretion governs the subsequent evolution of both the satellite and host halos (hereon, `satellite' and `host' refer to the lower and higher mass halo, respectively, during a merger).
Since galaxies form at the centers of dark matter halos \citep{WhiRee78,BluFabFlo86,Dub94,MoMaoWhi98}, the dynamics of infalling satellite halos will also influence the evolution of the accompanying galaxy populations.

After infall, a satellite halo can survive as a substructure halo (subhalo) of a larger host halo \citep{TorDiaSye98,KlyGotKra99,MooGhiGov99}.
As it orbits, a subhalo continues to host a (satellite) galaxy until it tidally disrupts or merges with the host halo's central galaxy\footnote{
Additionally, $\sim10\%$ of satellite galaxies merge with another satellite within the host halo \citep{AngLacBau09,WetCohWhi09a,WetCohWhi09b}.} \citep[e.g.,][]{SprWhiTor01,KraBerWec04,ZenBerBul05,ConWecKra06,NatKneSma09,WetWhi10}.
A satellite galaxy's survival timescale depends on its orbital parameters at the time of accretion, with galaxies on highly circular orbits surviving longer than those that rapidly plunge toward halo center.

Satellite orbits therefore influence the nature of galaxy mergers and the galaxy merger rate.
Galaxy mergers are expected to drive morphological evolution \citep{TooToo72,HauOst78,Whi78,BarHer96}, and the properties of the post-merger galaxy depend on the relative dynamics of the galaxies during the merger \citep{BoyMaQua05,CoxDutDiM06,HopHerCox08a}.
It is thus necessary to have cosmological predictions for how close satellite galaxies come to the central galaxy during first pericentric passage and the relative velocities distributions when they merge.

After infall, satellites also can experience mass loss from tidal stripping, tidal heating, and disk shocking \citep[e.g.,][]{OstSpiChe72,GneHerOst99,DekDevHet03,TayBab04,DOnSprHer10}.
Satellite galaxies are also thought to experience truncated star formation and/or morphological evolution, arising from ram-pressure stripping, adiabatic heating (strangulation), harassment, and/or tidal shock heating \citep[e.g.,][]{GunGot72,MooLakKat98,AbaMooBow99,McCFreFon08}.
Resonant stripping processes may also drive the evolution of dwarf spheroidals around the Milky Way \citep{DOnBesCox09}.
The efficiency of all of these processes depends critically on the details of satellite galaxy orbits.

Many semi-analytic models of satellite galaxy evolution contain prescriptions for dynamical friction survival times and tidal stripping \citep[see][for a recent review]{Bau06}.
A number of authors have provided detailed fits to satellite survival times that depend on a satellite's circularity at infall \citep{BoyMaQua08,JiaJinFal08}.
All these models require proper initial conditions of satellite orbits as a function of both halo mass and redshift to be an accurate depiction of galaxy evolution in a cosmological context.

Several authors have examined the orbital parameters of infalling satellite halos in simulations  \citep{Tor97,VitKlyKra02,Ben05,WanJinMao05,ZenBerBul05,KhoBur06}.
However, most work has focused on orbits only at $z\sim0$, and limited statistics and dynamic ranges have inhibited a robust investigation into possible mass and redshift dependence.
The nature of large-scale structure formation suggests that satellite accretion will depend on halo mass and redshift.
Triaxial collapse models of halo formation predict that more massive halos form from more spherical regions, with accreting matter containing less specific angular momentum \citep{Zel70,BBKS86,EisLoe95,SheMoTor01}.
This is reflected in the nature of the cosmic web, in which low-mass halos reside primarily along/within filaments while high-mass halos reside at the intersection of several filaments \citep{BonKofPog96}.
In this picture, matter infall onto massive halos occurs via narrow filaments \citep{ColWhiJen99,AubPicCol04,FalAllGot05} and is more radial than matter infall onto low-mass halos, which are comparable in size to their filament and thus experience more isotropic infall.

The nature of mass accretion may also vary with redshift, with implications for the formation of the earliest galaxies.
For instance, hydrodynamic simulations suggest that galaxy formation at $z\gtrsim2$ proceeds through highly radial flows of cold gas that penetrate deep into the host halo \citep{KerKatWei05,DekBirEng08}, behavior not observed in the local Universe.
While these radial flows are influenced by complex gas processes, the dynamics of matter at infall are governed primarily by gravity, so this trend suggests that mass accretion onto massive halos at high redshift is significantly more radial than at low redshift.

In this paper, we use a high-resolution $N$-body simulation of cosmological volume to examine the orbits of satellite halos at the time of infall into  larger host halos.
In addition to exploring the distributions of satellite orbital parameters, we also examine how the nature of satellite accretion varies with halo mass and redshift.
The combination of high resolution and large volume allows us to study the nature of accretion from dwarf galaxy masses ($10^{10}\hmsol$) to massive galaxy clusters ($10^{15}\hmsol$) with good statistics from $z=0$ to $5$.

\section{Methods} \label{sec:methods}
\subsection{Simulations \& Halo Tracking} \label{sec:simulations}

To find and track halos, we employed a dissipationless $N$-body simulation using the TreePM code of \citet{TreePM}.
This simulation used $\Lambda$CDM cosmology ($\Omega_{\rm m}=0.25$, $\Omega_{\rm \Lambda}=0.75$, $h=0.72$, $n=0.97$ and $\sigma_8=0.8$) in agreement with a wide array of observations \citep{DunKomNol08,KowRubAld08,VikKraBur09,PerReiEis10}.
For high mass and force resolution, the simulation evolved $1500^3$ particles in a $200\hmpc$ box, with a particle mass of $1.64 \times 10^8\hmsol$ and a Plummer equivalent smoothing of $3\hkpc$.
Initial conditions were generated using second-order Lagrangian Perturbation Theory at $z=250$ where the RMS was $20\%$ of the mean inter-particle spacing.
$45$ outputs were stored evenly in $\ln(a)$ from $z=10$ to $0$, with an output time spacing of $\sim650\,$Myr at $z=0$.
Note that the same simulation was used in \citet{WetWhi10}.

Halos were found using the Friends-of-Friends (FoF) algorithm \citep{DEFW85} with a linking length of $b=0.168$ times the mean inter-particle spacing.\footnote{
The longer linking length of $b=0.2$ is often used, but it is more susceptible to joining together distinct, unbound structures.}
We kept all FoF groups with more than $60$ particles, corresponding to $10^{10}\hmsol$.
This resolution level ensures that FoF halo masses are accurate to within $10\%$ \citep{WarAbaHol06}.
Merger trees were constructed from the set of halo catalogs by specifying a
parent-child relationship.
To be a `parent', a halo must contribute more than half of its mass to a `child' halo at the next simulation output.
Thus, a parent halo can never have more than one child.
A merger was then identified as a child halo with more than one parent.
Note that only distinct FoF halos (no subhalos) were used in this work.

\subsection{Ejected Halos \& Re-mergers} \label{sec:ejected}

Once a satellite halo has fallen into a larger halo, it can retain its identity as a bound subhalo.
In some cases a subhalo can become ejected from its host halo, either from being on an initially unbound orbit or from a scattering event within the host halo.
Up to $50\%$ of all satellite halos within $2-3\,\rvir$ of a host halo are a recently ejected population, particularly at low satellite halo mass (since subhalos experience severe mass stripping as they pass through a host halo) and for satellites with highly eccentric orbits \citep{GilKneGib05,LudNavSpr09,WanMoJin09}.
Including these re-merging satellite halos would bias the orbital distribution results both by double counting a single satellite halo across redshift and by artificially enhancing the population of highly elliptical orbits.

Since our merger tree does now allow a parent halo to have more than one child, recently ejected halos are necessarily parent-less.
Thus, we identified ejected halos (and their subsequent re-mergers) by finding halos which have no parent halo but had at least one of their 20 most bound particles within some halo at the previous output.\footnote{
We checked that this method gives consistent results as compared with the subhalo catalogs of \citet{WetWhi10}.}
We define infall as the first time a halo merges with a more massive halo, and we used this event to compute satellite orbits, discarding subsequent re-mergers.
A re-merger was not discarded, however, if the satellite halo had more than doubled its mass since becoming ejected, since in this case we considered a new halo to have formed.

\subsection{Calculating Orbits} \label{sec:calcorbit}

To calculate the orbital parameters of satellite halos at infall into a more massive host halo, we used the simulation halo merger trees to find halos which were about to merge, that is, become joined within an isodensity contour by the FoF algorithm.
We then computed the orbital parameters using halo masses, positions, and velocities in the output prior to merging.
Positions and velocities were those of the halo's most bound particle, which is expected to correspond to a hosted galaxy.
Velocities were calculated using physical coordinates, including Hubble flow.

With these halo properties, we calculated orbital parameters by treating the two halos as isolated point particles in the reduced mass frame (limitations of this approximation are discussed below and in the Appendix).
Given satellite mass $\ms$ and host halo mass $\mh$, with reduced mass $\mu = \ms\mh/\left(\ms+\mh\right)$, satellite separation $\mathbf{r} = \mathbf{r}_{\rm host} - \mathbf{r}_{\rm sat}$ and velocity $\dot{\mathbf{r}} = \dot{\mathbf{r}}_{\rm host} - \dot{\mathbf{r}}_{\rm sat}$, the orbital energy is
\begin{equation}
E = \frac{1}{2}\mu\dot{r}^2 - \frac{G\ms\mh}{r}
\end{equation}
and the angular momentum is
\begin{equation}
L = \mu \mathbf{r} \times \mathbf{\dot{r}}\, .
\end{equation}
With these, we computed the orbital eccentricity,
\begin{equation}
e = \sqrt{1+\frac{2EL^2}{\left(G\ms\mh\right)^2\mu}}
\end{equation}
pericentric distance (from host halo center),
\begin{equation} \label{eq:rperi}
\rperi = \frac{L^2}{(1+e)G\ms\mh\mu}
\end{equation}
and apocentric distance
\begin{equation}
\rapo = \frac{L^2}{(1-e)G\ms\mh\mu} \, .
\end{equation}
Orbital circularity is defined as the ratio of the orbit angular momentum to that of the circular orbit with the same energy and can also be related to eccentricity
\begin{equation}
\eta = \frac{j(E)}{j_{\rm c}(E)} = \sqrt{1-e^2} \, .
\end{equation}

In the point particle two-body approximation, the above quantities are all conserved throughout the orbit, so they do not depend on the separation of the satellite from the host halo.
Two of the above quantities (for example, circularity plus pericenter) are sufficient to uniquely describe an orbit.
An alternate description can be given by the satellite's radial, $\vr$,  plus tangential, $\vt$, velocity, though they need to be defined at a given radius, which here will be at the host halo virial radius.
To do this, we used the conservation of orbital energy and angular momentum to evolve the velocities from the satellite's measured location to where its center crosses the host halo virial radius, $\rvir$, which was derived from the halo FoF mass and concentration assuming a spherical NFW \citep{NFW96} density profile.

Hereon, all satellite halo distances and velocities are scaled to the host halo virial radius and virial circular velocity, $\vvir = \sqrt{G\mvir/\rvir}$.

\subsection{Calculating Orbital Distributions} \label{sec:calcdistribution}

The aim of this work is to examine the orbital parameter distributions of satellite halos as they first pass through the virial radius of a larger host halo.
However, because of the finite time spacing of simulation outputs, satellite halos were identified at a variety of distances from the host halo virial radius prior to merging.
This presents some difficulty in directly measuring orbital parameter distributions, since satellite halos can experience mass loss (or growth), dynamical friction, and tidal forces, all of which alter the orbital parameters as satellites traverse significant distances.

To circumvent this issue of finite time resolution, we estimated satellite orbital parameter distributions by selecting only satellite halos that were within a small separation of the host halo virial radius at the output prior to merging \citep[as done by][]{VitKlyKra02,Ben05,WanJinMao05}.
Specifically, we found satellite halos that were about to merge with a larger host halo and whose edge (given by its own virial radius) was in range $[1,1+\Delta r]\,\rvir$ of the virial radius of the host halo.\footnote{
We used the location of the satellite edge since the FoF algorithm merges halos when the satellite edge crosses the host halo $\rvir$.}
We used $\Delta r = 0.25$, finding no significant difference using smaller values.

Since more radial orbits spend less time within the given radial shell, to properly estimate the overall orbital distributions one must scale the orbital counts by the crossing time within the shell.
Specifically, we scaled the counts by $\Delta t/t_{\rm cross}$, where $\Delta t$ is the output time interval and $t_{\rm cross}$ is the orbital crossing time for a satellite halo edge to orbit from a separation of $\Delta r$ to $0$.\footnote{
This correction for crossing time was employed by \citet{Ben05} and \citet{WanJinMao05} but was neglected in \citet{VitKlyKra02}.}

Since this method of estimating satellite orbital distributions considers only satellites close to the location of interest, it is little-affected by mass stripping, dynamical friction, and tidal forces, and the error on the estimated orbital parameters is expected to be less than $10\%$ \citep{Ben05}.
There are, however, limitations to the point particle two-body approximation, which ignores multi-body interactions and extended, triaxial halo profiles (see the Appendix).
Some satellites (up to $20\%$ at $z=0$) had their $\rperi$ or $\rapo$ within the radial selection shell, so these objects did not have a well-defined crossing time even though they did merge with the host halo in the next output.
This population was dominated by satellites on highly circular orbits, such that
$\rapo$ was never beyond the outer radial shell.
In weighting their counts for orbital distributions, these satellites were given a shell crossing time equal to the longest well-defined crossing time satellite in the given output, which amounts to a minimal weighting without discarding them.
Additionally, some merging satellites were on formally unbound orbits, though this represents less than $2\%$ of satellites regardless of mass or redshift.\footnote{
Another method to estimate satellite orbital parameter distributions is to consider all satellite halos which are about to merge, regardless of their distance from the host halo in the output prior to merging, as done by \citet{KhoBur06}.
However, they found that in order to select systems which conserve energy and angular momentum, they were restricted to `isolated' merger systems which do not change in total mass by more than $10\%$ between outputs.
Indeed, we found that using all satellites which are about to merge yields as high as $15\%$ of orbits being formally unbound, since satellite halos on highly eccentric orbits, which are more likely to be found at large radii in the output prior to merging, become significantly more bound prior crossing the host halo virial radius.
We found that while using this alternate method yielded quantitatively different orbital distributions, it yielded qualitatively similar results regarding the orbital distribution shapes and their mass and redshift dependence.}

Finally, we also examined how using $M_{200{\rm crit}}$ instead of FoF mass influences the results.\footnote{
The relation of FoF($b=0.168$) mass to spherical overdensity ($200{\rm crit}$) mass depends on halo concentration and redshift, though for the regimes considered here they are within $\sim15\%$.}
We found that the averages of the orbital distributions from the two methods are within error, and thus the specific halo finding algorithm used does not significantly affect the results, in agreement with similar tests of \citet{Ben05}.

\section{Orbital Distributions at $z=0$} \label{sec:distribution}

\begin{figure}
\begin{center}
\resizebox{3in}{!}{\includegraphics{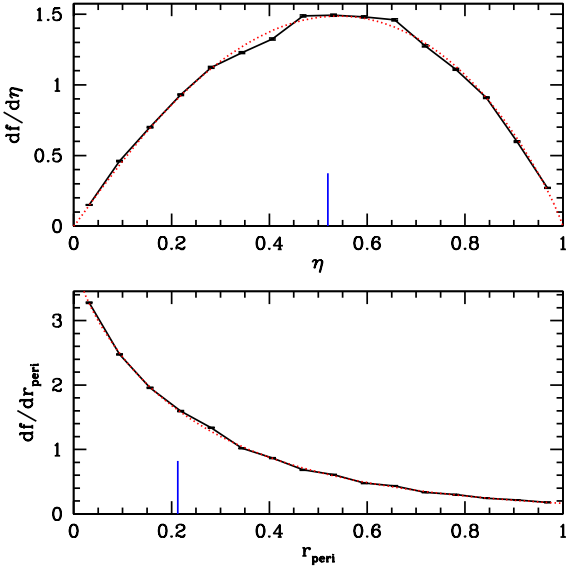}}
\end{center}
\vspace{-0.1in}
\caption{
Distributions of circularity, $\eta$, and pericenter, $\rperi$ (scaled to host halo virial radius), of infalling satellite halos at $z=0$, for halos $>10^{10}\hmsol$.
Errors bars indicate Poisson error in each bin.
Vertical lines show average circularity and median pericenter.
Dotted curves show fits to the distributions (Eqs.~\ref{eq:circfit} and \ref{eq:rperifit}).
} \label{fig:circ_rperi_dist}
\end{figure}

\begin{figure}
\begin{center}
\resizebox{3in}{!}{\includegraphics{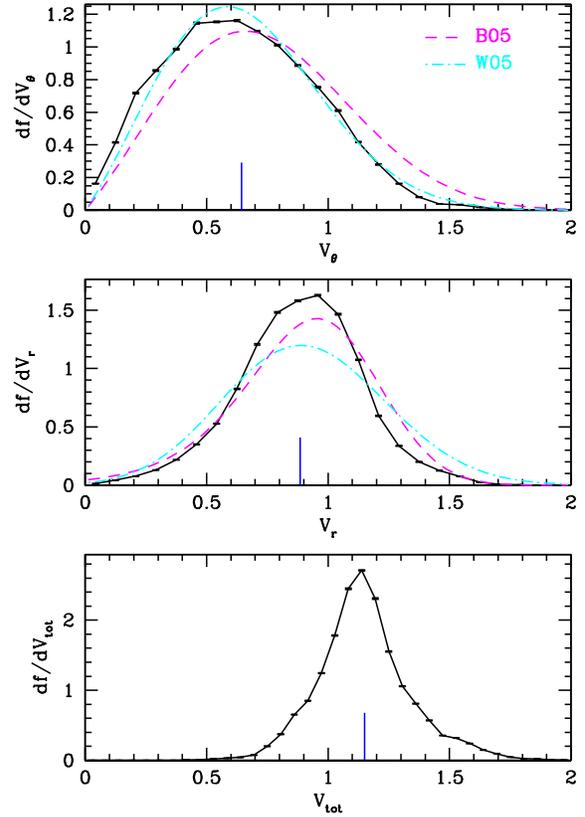}}
\end{center}
\vspace{-0.1in}
\caption{
Distributions of tangential, $\vt$, radial, $\vr$, and total, $\vtot$, velocity (scaled to host halo virial velocity) of infalling satellite halos at $z=0$, for halos $>10^{10}\hmsol$.
Vertical lines show average velocities.
Dashed and dot-dashed curves show distributions from \citet{Ben05} and \citet{WanJinMao05}.
} \label{fig:vt_vr_vtot_dist}
\end{figure}

We begin by examining the distributions of satellite orbital parameters for all infalling satellites at $z=0$.
Figure~\ref{fig:circ_rperi_dist} shows the distribution of circularity and pericenter, scaled to the host halo virial radius, for all halos $>10^{10}\hmsol$.
Circularity shows a broad distribution, peaking at $\eta=0.52$ (which corresponds to an eccentricity of $e=0.85$), in good agreement with earlier work \citep{Tor97,ZenBerBul05,WanJinMao05,KhoBur06}.
Satellite orbits tend to be neither highly radial nor highly circular.
Pericenter shows a distribution falling exponentially with radius, with a median value of $\rperi=0.21$ and falling to zero for $\rperi>1$.
Most infalling satellites are on orbits taking them close to the center of the host halo.

While an orbit can be fully classified by its circularity and pericenter, another such combination is radial and tangential velocity as measured at the host halo virial radius.
Figure~\ref{fig:vt_vr_vtot_dist} (top and middle) shows these satellite velocity distributions, scaled to the host halo virial velocity.
Since the velocities are computed at the host halo virial radius, the tangential velocity is equivalent to the satellite specific angular momentum: $L/L_{\rm vir}=\vt/\vvir$.
Most satellites infall with less specific angular momentum than the virial value of the host halo, with an average $\vt=0.64$.
However, this does not mean that satellite mergers do not contribute significantly to the angular momentum build-up of halos, since the actual spin of a halo is only $\approx5\%$ of its virial value \citep[e.g.,][]{VitKlyKra02}.
Indeed, the merging of satellites is the primarily driver of a halo's angular momentum \citep{VitKlyKra02}.

As opposed to satellite tangential velocity, radial velocity is typically comparable to the host halo virial velocity, with an average $\vr=0.89$.
Taken together, these imply that most satellites are infalling `hotter' than their host halo.
Figure~\ref{fig:vt_vr_vtot_dist} (bottom) demonstrates this explicitly, showing the satellite total velocity distribution, again scaled to the host halo virial velocity.
On average, satellites are infalling with $\sim15\%$ higher velocity than the matter within the host halo.

To compare with previous work, Fig.~\ref{fig:vt_vr_vtot_dist} also shows fits to the velocity distributions at $z=0$ of \citet{Ben05} and \citet{WanJinMao05}, which represent a Gaussian distribution for radial velocity and approximately a two-dimensional Maxwell-Boltzmann distribution for tangential velocity.
While the radial velocity distribution seen here exhibits a somewhat narrower profile, the distributions show overall broad agreement, particularly in the averages and peak locations.
Some of the differences may be attributed to previous works using somewhat different cosmology and that \citet{WanJinMao05} selected satellite \textit{subhalos} (halos which may have already merged) within a radial shell.
More importantly, the previous works examined satellite orbits at a higher mass regime, and as will be explored in the next section, satellite orbital distributions depend on halo mass.

Finally, the above tangential velocity distribution explains why pericenter exhibits an exponential distribution.
At fixed mass ratio, $\rperi \propto \vt^2/(1+e)$ (Eq.~\ref{eq:rperi}).
Since eccentricity is essentially limited to $0<e<1$, the $\rperi$ distribution is dominated by the $\vt^2$ term.
$\vt$ is described by a two-dimensional Maxwell-Boltzmann distribution, $P(\vt) \sim \vt {\rm e}^{-(\vt-V_o)^2}$, so under transformation, the distribution of $\vt^2$, and hence $\rperi$, is exponential.
Note that considering orbits in extended host halo profiles will modify this exponential distribution somewhat (see the Appendix).

\section{Mass Dependence} \label{sec:mass}

We next explore how satellite orbital distributions depend on halo mass at $z=0$.
We first examined the dependence on host halo mass by selecting satellite halos in the mass range $10^{10.0-10.5}\hmsol$, and we used the large dynamic range of the simulation to explore the dependence across four decades in halo mass.

\begin{figure}
\begin{center}
\resizebox{3in}{!}{\includegraphics{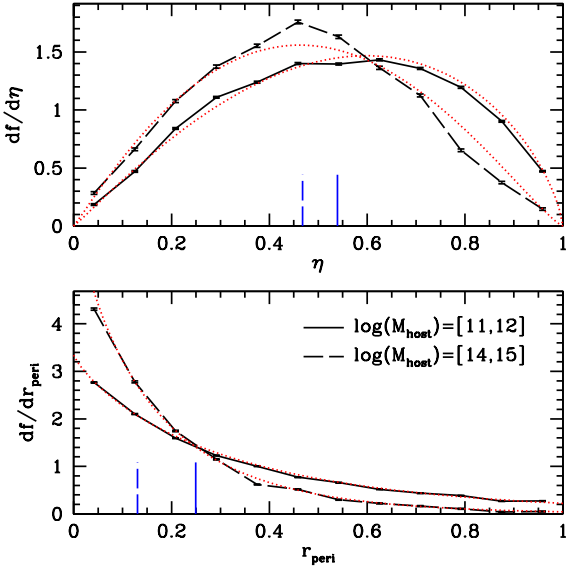}}
\end{center}
\vspace{-0.1in}
\caption{
Distributions of circularity and pericenter at $z=0$, for satellite halos of mass $10^{10.0-10.5}\hmsol$ and host halos in two mass ranges.
Vertical lines show average circularity and median pericenter.
Dotted curves show fits to the distributions (Eqs.~\ref{eq:circfit} and \ref{eq:rperifit}).
Satellite orbital distribution shapes depend on host halo mass.
} \label{fig:circ_rperi_dist_mhost}
\end{figure}
 
Figure~\ref{fig:circ_rperi_dist_mhost} shows the distribution of circularity and pericenter, as in Fig.~\ref{fig:circ_rperi_dist}, but for two host halo mass ranges.
The distributions clearly change shape, being skewed to both lower circularity and lower pericenter for more massive host halos, which drives the average/median of the distributions (vertical lines) down.

\begin{figure}
\begin{center}
\resizebox{3in}{!}{\includegraphics{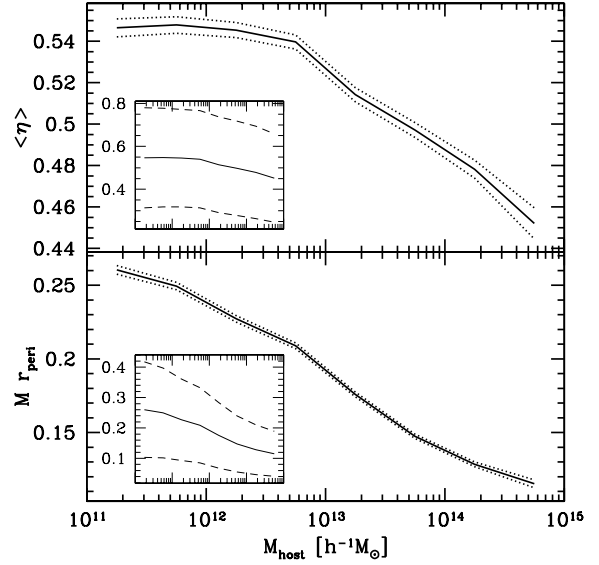}}
\end{center}
\vspace{-0.1in}
\caption{
Dependence of satellite average circularity and median pericenter on host halo mass at $z=0$, for satellite halos of mass $10^{10.0-10.5}\hmsol$.
Dotted curves show standard errors of the mean/median.
Inset panels show same, but dashed curves indicate the standard deviation (or $68$ percentile width) of the full distribution.
Satellite orbits are both more radial and plunge deeper at higher host halo mass.
} \label{fig:circ_rperi_mhost}
\end{figure}

Figure~\ref{fig:circ_rperi_mhost} demonstrates explicitly the dependence of average circularity and median pericenter on host halo mass.
Circularity shows no dependence up to $\sim3\times10^{12}\hmsol$, but above this mass satellite orbits become less circular with increasing host halo mass, with average circularity dropping nearly $20\%$ across the mass range.
Interestingly, the turnover corresponds to the value of $M_*$, the characteristic halo mass scale of collapse, at $z=0$.\footnote{
More specifically, $M_*(z)$ is the mass at which $\sigma(M,z)$, the variance of the linear power spectrum at redshift $z$ smoothed on scale $M$, equals the threshold for linear density collapse, $\delta_{\rm c}=1.69$.}
Median pericenter decreases more strongly with host halo mass, falling by more than a factor of 2 across the mass range and showing no rollover at low mass.
Overall, satellite orbits are both more radial and plunge deeper into their host halo at higher host halo mass.

\begin{figure}
\begin{center}
\resizebox{3in}{!}{\includegraphics{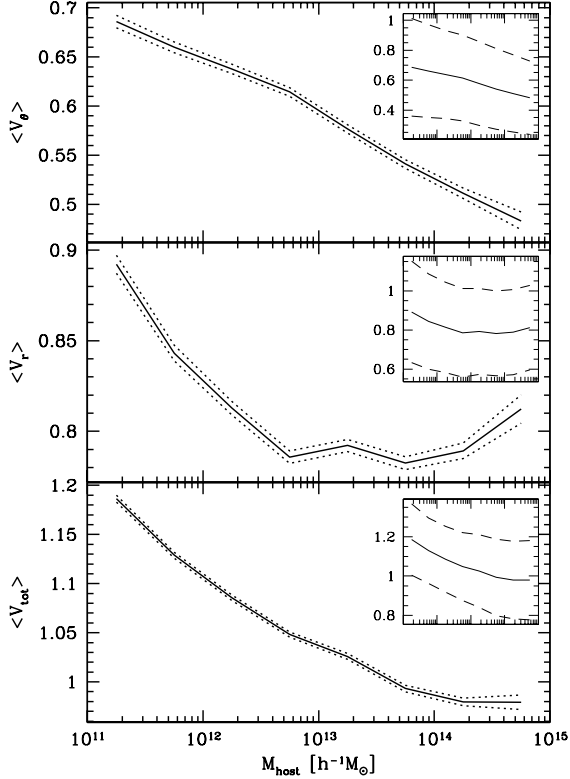}}
\end{center}
\vspace{-0.1in}
\caption{
Dependence of satellite average tangential, radial, and total velocity on host halo mass at $z=0$, for satellite halos of mass $10^{10.0-10.5}\hmsol$.
Inset panels show same, but dashed curves indicate the standard deviation of the full distribution.
The degree to which satellite infall velocities are biased relative to their host halos depends on host halo mass.
} \label{fig:vt_vr_vtot_mhost}
\end{figure}

Similarly, Fig.~\ref{fig:vt_vr_vtot_mhost} (top and middle) shows the host halo mass dependence of satellite average tangential and radial velocity.
Both velocity components remain below the host halo virial velocity at all mass scales.
Tangential velocity monotonically declines with host halo mass, falling by $30\%$, implying that satellite accretion contributes angular momentum less efficiently onto higher mass halo.
Radial velocity declines rapidly with host halo mass up to $\sim3\times10^{12}\hmsol$, beyond which it remains nearly flat.

These trends explain the host halo mass dependence of circularity and pericenter.
Below $\sim3\times10^{12}\hmsol$, both $\vt$ and $\vr$ decline with host halo mass, giving rise to constant circularity.
At fixed circularity (eccentricity) and $\ms$, $\rperi/\rvir \propto \vt^2/\mh^{4/3}$ (Eq.~\ref{eq:rperi}), so $\rperi$ continues to decline rapidly with mass.
Above $\sim3\times10^{12}\hmsol$, declining $\vt$ and constant $\vr$ cause both circularity and pericenter to decline with mass.

Additionally, Fig.~\ref{fig:vt_vr_vtot_mhost} (bottom) shows the host halo mass dependence of satellite average total velocity.
While satellite infall is usually `hotter' than the host halo virial velocity, the magnitude of this effect decreases with host halo mass, falling by $20\%$ over the mass range.
Interestingly, this leads to a crossover such that satellites infalling onto halos $>10^{14}\hmsol$ are instead on average `colder' than the host halo.
This mass trend is driven by the interplay of halo mass with its environment.
High-mass halos generically dominate the local potential field, and so satellites naturally fall in with velocity comparable to the halo virial velocity.
By contrast, many low-mass halos reside in proximity to higher mass halos which significantly boost the local potential field, causing satellites to fall into low-mass halos faster than their virial velocity \citep[see also Fig.~8 in][on how mass, separation, and environment affect halo mergers]{WetSchHol08}.

To highlight the significance of these mass dependencies with respect to the overall orbital distributions, the inset panels of Figs.~\ref{fig:circ_rperi_mhost} and \ref{fig:vt_vr_vtot_mhost} show the same as the main panels, but the dashed curves indicate the full distribution width (standard deviation).
While mass dependence is significant in each case, note that the mean/median  changes by at most one standard deviation across the mass range probed here.

\begin{figure}
\begin{center}
\resizebox{3in}{!}{\includegraphics{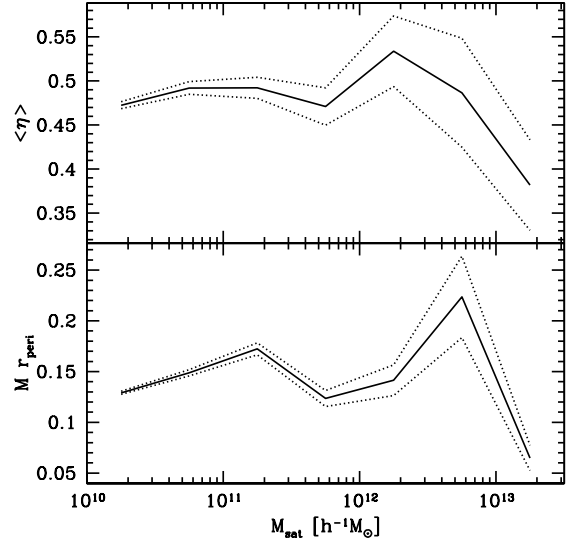}}
\end{center}
\vspace{-0.1in}
\caption{
Dependence of satellite average circularity and median pericenter on satellite halo mass at $z=0$, for host halos of mass $10^{14-15}\hmsol$.
No significant, systematic dependence on satellite mass is found at any mass scale or redshift.
} \label{fig:circ_rperi_msat}
\end{figure}

While the above results demonstrate that satellite orbital parameters depend on host halo mass, Fig.~\ref{fig:circ_rperi_msat} explores whether circularity and pericenter depend on satellite halo mass, for host halos of fixed mass $10^{14-15}\hmsol$.
While Fig.~\ref{fig:circ_rperi_msat} suggests a possible drop in circularity and pericenter for very massive satellites, the large scatter across satellite halo mass does not lead to any clear, systematic trends.
Moreover, we found no clear trends with satellite mass for \textit{any} orbital parameters, including varying host halo mass or redshift.
Thus, we conclude that the nature of satellite infall is controlled by host halo mass and is little affected by satellite mass.
This supports the physical picture that orbital dynamics are governed by the most massive halo within a region, and that less massive satellites effectively act as massless tracers of the potential field.

\section{Redshift Evolution} \label{sec:evolution}

We now turn to explore the dependence of satellite orbital parameters on redshift.
Since the halo mass function declines significantly with redshift, examining the mergers of all halos above the resolution limit ($>10^{10}\hmsol$) would convolve any redshift dependence together with the mass dependence of the previous section.
So to isolate redshift trends, we examined halos in a fixed mass range: host halos of mass $10^{12.0-12.5}\hmsol$ (corresponding to $M_*$ halos at $z=0$) and satellite halos of mass $10^{10-11}\hmsol$.

\begin{figure}
\begin{center}
\resizebox{3in}{!}{\includegraphics{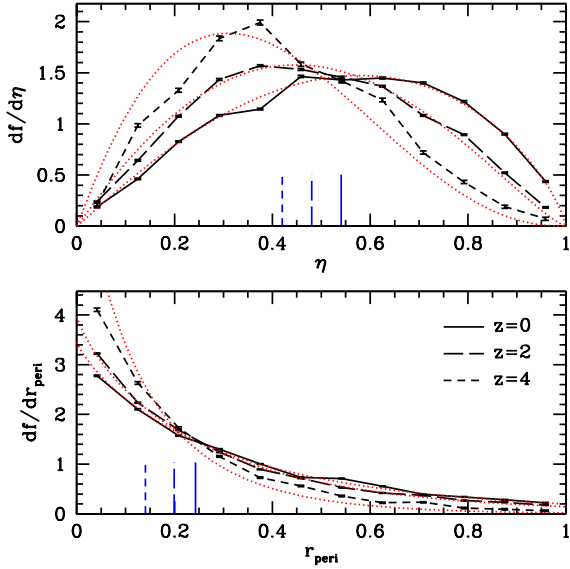}}
\end{center}
\vspace{-0.1in}
\caption{
Distributions of circularity and pericenter at different redshifts, for host halos of mass $10^{12.0-12.5}\hmsol$ and satellite halos of mass $10^{10-11}\hmsol$.
Dotted curves show fits to the distributions (Eqs.~\ref{eq:circfit} and \ref{eq:rperifit}).
Satellite orbital distribution shapes depend on redshift, though the distribution widths remain broad at all redshifts. 
} \label{fig:circ_rperi_dist_evol}
\end{figure}

\begin{figure}
\begin{center}
\resizebox{3in}{!}{\includegraphics{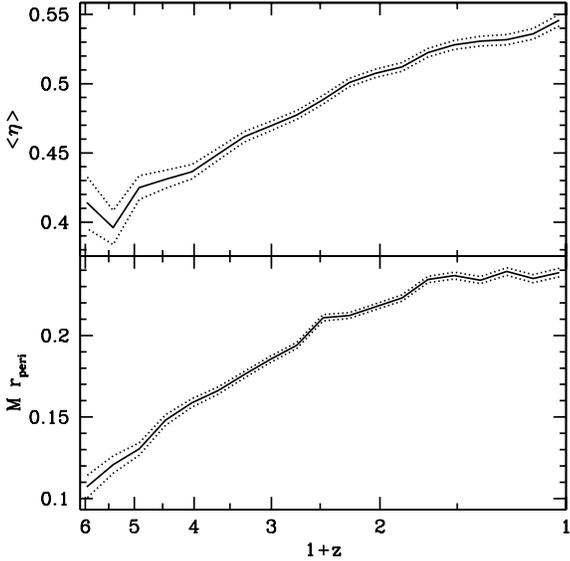}}
\end{center}
\vspace{-0.1in}
\caption{
Dependence of satellite average circularity and median pericenter on redshift, for host halos of mass $10^{12.0-12.5}\hmsol$ and satellite halos of mass $10^{10-11}\hmsol$.
Satellite orbits are both more radial and plunge deeper at higher redshift.
} \label{fig:circ_rperi_evol}
\end{figure}

Figure~\ref{fig:circ_rperi_dist_evol} shows the distribution of circularity and pericenter at three redshifts.
Similar to the dependence on host halo mass in Fig.~\ref{fig:circ_rperi_dist_mhost}, the distributions are skewed to both lower circularity and lower pericenter at higher redshift.
Figure~\ref{fig:circ_rperi_evol} demonstrates more explicitly the dependence of average circularity and median pericenter on redshift.
Average circularity falls nearly $30\%$ from $z=0$ to $5$, while pericenter falls more rapidly to less than half of its $z=0$ value.
Overall, satellite orbits become more radial and plunge deeper into their host halos with increasing redshift.

\begin{figure}
\begin{center}
\resizebox{3in}{!}{\includegraphics{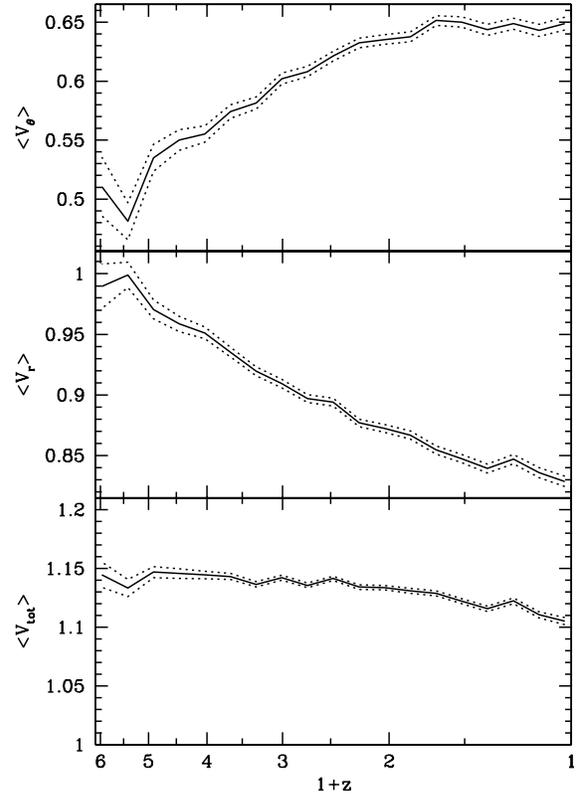}}
\end{center}
\vspace{-0.1in}
\caption{
Dependence of satellite average tangential, radial, and total velocity on redshift, for host halos of mass $10^{12.0-12.5}\hmsol$ and satellite halos of mass $10^{10-11}\hmsol$.
Tangential velocity declines with redshift, while radial velocity increases, leading to nearly constant total velocity.
} \label{fig:vt_vr_vtot_evol}
\end{figure}

Figure~\ref{fig:vt_vr_vtot_evol} also shows the redshift evolution of satellite velocities.
Tangential velocity declines with redshift, implying that satellite accretion less efficiently contributes angular momentum growth at higher redshift.
Conversely, radial velocity increases with redshift, approaching the host halo virial value at $z\approx5$.
Interestingly, the contrasting evolutionary trends of radial and tangential velocity lead to no significant evolution in satellite total velocity, which at all redshifts remains `hotter' than the host halo virial velocity.

Note that the widths of the total distributions do not evolve significantly with redshift, retaining standard deviations of $\approx 0.2$ as shown in Fig.~\ref{fig:circ_rperi_dist_evol}.
Thus, as with mass dependence, the redshift evolution of the mean/median out to $z=5$ is comparable to the distribution width itself.

\begin{figure}
\begin{center}
\resizebox{3in}{!}{\includegraphics{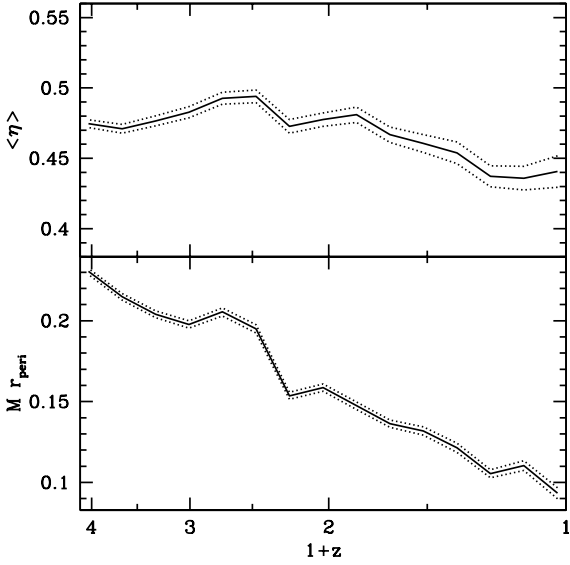}}
\end{center}
\vspace{-0.1in}
\caption{
Same as Fig.~\ref{fig:circ_rperi_evol}, but for host halos of mass $270\,M_*(z)$ ($10^{14.75}\hmsol$ at $z=0$ down to $10^{10.3}\hmsol$ at $z=3$) and satellite halos of mass $10^{10.0-10.3}\hmsol$.
When scaled by $M_*(z)$, circularity remains nearly constant while pericenter increases significantly.
} \label{fig:circ_rperi_evol_mstar}
\end{figure}

What drives the redshift evolution of satellite orbits?
One possibility is that redshift dependence is simply a manifestation of the trends with mass from the previous section.
Analytical triaxial collapse models of halo formation \citep{BBKS86,EisLoe95,SheMoTor01} predict a self-similarity in the nature of matter infall with redshift at fixed $M/M_*(z)$, since it is at $M_*$ that halos transition from being located along filaments to being at the intersection of several filaments.
In this picture, the above redshift dependence is driven by probing higher $\mh/M_*(z)$ at higher $z$.

To investigate this, we selected host halos of mass $\mh/M_*(z)=270$ at each redshift (corresponding to $10^{14.75}\hmsol$ at $z=0$ down to $10^{10.3}\hmsol$ at $z=3$) and satellite halos in the range $10^{10.0-10.3}\hmsol$.
Figure~\ref{fig:circ_rperi_evol_mstar} (top) shows that satellite infall into host halos of constant $\mh/M_*(z)$ indeed exhibits nearly constant circularity.
However, pericenter (bottom) is not constant but instead increases with redshift by a factor of 2 (opposite to the trend at fixed mass).
Furthermore, at fixed $\mh/M_*(z)$ we found a significant increase in radial, tangential, and total velocity with redshift.
Thus, while the redshift evolution can be partially understood simply as a manifestation of mass dependence, intrinsic redshift dependence does exist.

\section{Fits to Orbital Distributions} \label{sec:fit}

\begin{figure}
\begin{center}
\resizebox{3in}{!}{\includegraphics{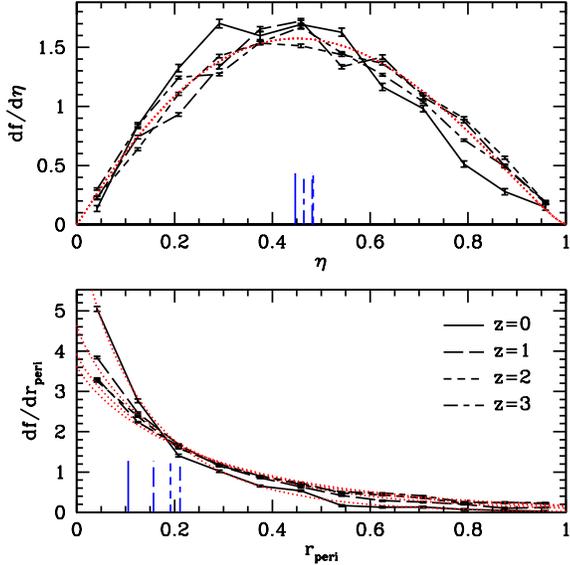}}
\end{center}
\vspace{-0.1in}
\caption{
Distributions of circularity and pericenter at different redshifts, for host halos of mass $270\,M_*(z)$ and satellite halos of mass $10^{10.0-10.3}\hmsol$.
Dotted curves show fits to the distributions (Eqs.~\ref{eq:circfit} and \ref{eq:rperifit}).
The circularity distribution is nearly universal for host halos of fixed $\mh/M_*(z)$.
} \label{fig:circ_rperi_dist_evol_mstar}
\end{figure}

We now seek to provide analytical fits to the orbital distributions of the two parameters of primary interest, circularity and pericenter.
The results of the previous sections show that the shapes of these orbital distributions depend sensitively on both mass and redshift.
However, Fig.~\ref{fig:circ_rperi_evol_mstar} shows that, at fixed $\mh/M_*(z)$, circularity remains nearly constant with redshift.
Figure~\ref{fig:circ_rperi_dist_evol_mstar} (top) demonstrates this more explicitly, showing that the overall circularity distribution remains approximately universal at fixed $\mh/M_*(z)$.
However, the distribution of pericenter is neither redshift-invariant at fixed $\mh$ (Fig.~\ref{fig:circ_rperi_dist_evol}) nor at fixed $\mh/M_*(z)$ (Fig.~\ref{fig:circ_rperi_dist_evol_mstar}).

\begin{table}
\begin{center}
\begin{tabular}{|c|cccc|}
\hline
& $\alpha_i$ & $\beta_i$ & $\gamma_i$ & $ g_i(z) $ \\
\hline
$C_0$ & $3.38$ & $0.567$ & $0.152$ & $1$ \\
$C_1$ & $0.242$ & $2.36$ & $0.108$ & $1$ \\
\hline
$R_0$ & $3.14$ & $0.152$ & $0.410$ & $(1+z)^{-4}$ \\
$R_1$ & $0.450$ & $-0.395$ & $0.109$ & $(1+z)^{-4}$ \\
\hline
\end{tabular}
\end{center}
\caption{Fit parameters for $C_i,R_i$, given by Eq.~\ref{eq:fitparams}, for use in circularity (Eq.~\ref{eq:circfit}) and pericenter (Eq.~\ref{eq:rperifit}) distributions.}
\label{tab:circ_rperi_fit}
\end{table}

The orbital distributions of satellite circularity, $\eta$, and pericenter, $\rperi$, are given to good approximation across mass ($10^{10-15}\hmsol$) and redshift ($z=0-5$) by the following simple functional forms
\begin{equation} \label{eq:circfit}
\frac{{\rm d}f}{{\rm d}\eta} = C_0(\mh,z)\eta^{1.05}(1-\eta)^{C_1(\mh,z)}
\end{equation}
\begin{equation} \label{eq:rperifit}
\frac{{\rm d}f}{{\rm d}\rperi} = R_0(\mh,z){\rm exp}\left\{-[\rperi/{R_1(\mh,z)}]^{0.85}\right\} \, .
\end{equation}
Parameters $C_i(\mh,z)$ and $R_i(\mh,z)$ describe the mass and redshift dependence and take a similar functional form
\begin{equation} \label{eq:fitparams}
C_i,R_i = \alpha_i\left(1+\beta_i\left[g_i(z)\frac{\mh}{M_*(z)}\right]^{\gamma_i}\right)
\end{equation}
where the values for $\alpha_i,\beta_i,\gamma_i$ and the function $g_i(z)$ are given in Table~\ref{tab:circ_rperi_fit}.
Since the circularity distribution is nearly constant when scaled by $\mh/M_*(z)$, $g_i(z)=1$ for $C_i$.
While the pericenter distribution does not exhibit such redshift invariance, the scaling $g_i(z)\mh/M_*(z)$ with $g_i(z)=(1+z)^{-4}$ for $R_i$ fully encapsulates the redshift dependence.

Since parameters $C_0(\mh,z)$ and $R_0(\mh,z)$ are merely normalizations for probability distributions, their mass and redshift dependencies are implicitly given by those of $C_1(\mh,z)$ and $R_1(\mh,z)$.
We include fits to their mass and redshift dependencies for completeness.

Finally, given the assumed cosmology the evolution of $M_*(z)$ is approximated to within $5\%$ up to $z=7$ by
\begin{equation}
\log\left[M_*(z)/\hmsol\right] = 12.42 - 1.56z + 0.038z^2 \, .
\end{equation}

The orbital distribution fits given by Eqs.~\ref{eq:circfit} and \ref{eq:rperifit} provide a good description of both the mass and redshift dependence of satellite orbits.
The accuracy of these fits is demonstrated explicitly in Figs.~\ref{fig:circ_rperi_dist}, \ref{fig:circ_rperi_dist_mhost}, \ref{fig:circ_rperi_dist_evol}, and \ref{fig:circ_rperi_dist_evol_mstar}.

We note two caveats to these orbital parameter fits.
First, these are valid for satellites at the time of infall and were derived assuming point particle orbits.
Thus, they are likely to be modified well after infall as satellites experience extended, triaxial halo potentials (see the Appendix), as well as dynamical friction, mass stripping, and scattering processes.
Second, while we did not examine in detail correlations between the circularity and pericenter distributions, some correlation is expected, with more circular orbits tending to have larger pericenters, though with significant scatter \citep{Tor97,GilKneGib04,Ben05,KhoBur06}.
We leave a more detailed investigation into these behaviors to future work.

\section{Summary \& Discussion} \label{sec:summary}

We used a high-resolution $N$-body simulation of cosmological volume to track the mergers of halos from dwarf galaxy masses ($10^{10}\hmsol$) to massive galaxy clusters ($10^{15}\hmsol$) across a large redshift range ($z=0$ to $5$).
We explored the orbital parameters of satellite halos at infall, when they cross within the virial radius of a larger host halo.
The main results are as follows:
\begin{itemize}
\item Satellite orbital parameters exhibit broad distributions, with typical standard deviations of $\approx 0.2$.
For all resolved halos at $z=0$, the average values of the distributions are: circularity $\eta=0.52$, (median) pericenter $\rperi/\rvir=0.21$, tangential velocity $\vt/\vvir=0.64$, radial velocity $\vr/\vvir=0.89$, and total velocity $\vtot/\vvir=1.15$.
\item Satellite orbits are more radial and plunge deeper into their host halos at higher host halo mass, but orbits are not significantly affected by satellite halo mass.
\item Infalling satellites are typically `hotter' than the host halo virial velocity, except onto the most massive host halos where satellite are slightly `colder'.
\item At fixed halo mass, satellite orbits become more radial and plunge deeper into their host halos at higher redshift.
\item The satellite circularity distribution exhibits almost no redshift evolution for host halos of fixed $\mh/M_*(z)$, implying that redshift dependence of circularity at fixed mass is simply derivative of varying $M/M_*(z)$.
However, pericenter exhibits significant increase with redshift at fixed $\mh/M_*$.
\end{itemize}

As explored in \S\ref{sec:distribution}, the orbital distributions we find at $z=0$ when stacking halos of all masses agree well with those of previous work.
However, previous results of possible mass and/or redshift dependence of satellite orbits are mixed.
\citet{Ben05} found evidence that satellite orbits become more radial at higher halo mass scales but was unable to quantify this further, while \citet{VitKlyKra02} and \citet{WanJinMao05} saw no such halo mass dependence over a limited mass range.
By contrast, \citet{VitKlyKra02} found that satellite angular momentum decreases with increasing satellite mass to host halo mass ratio, while \citet{WanJinMao05} and \citet{KhoBur06} found no dependence on mass ratio, though again over a limited mass ratio range.
Finally, the results of \citet{Ben05} suggested trends with redshift while those of \citet{VitKlyKra02} did not.
In most cases, previous work was limited in terms of merger statistics (in some cases, examining infall into a handful of halos) and dynamical range (unable to explore both $M \gg M_*$ and $M \ll M_*$).
The results here demonstrate clear dependence of satellite orbits on host halo mass and redshift and no significant evidence for dependence on satellite mass.
This is broadly consistent with the predictions of analytical triaxial collapse models \citep{BBKS86,EisLoe95,SheMoTor01}, in which more massive halos arise from a more spherical Lagrangian volume with less angular momentum.
We emphasize that the existence of mass and redshift dependence implies that fits to orbital distributions based on stacking halos of all masses at $z\sim0$ \citep[e.g.,][]{Ben05,ZenBerBul05,WanJinMao05} are not universally accurate.

The mass and redshift trends seen here have implications for various aspects of galaxy formation and evolution.
For example, recent work suggests that galaxy formation at $z \gtrsim 2$ proceeds through narrow streams of cold gas \citep{KerKatWei05,DekBirEng08}, fundamentally different behavior than seen in the local Universe.
\citet{DekBir06} argued that this difference arises in part because massive ($\sim10^{12}\hmsol$) halos at high redshift form at the intersection of narrow filaments, while at low redshift such halos are more likely embedded within a filament and experience wide-angle inflow.
Our results on the redshift evolution of satellite accretion qualitatively support this picture, but it is not clear that the evolution seen here is strong enough to imply a fundamental change the nature of accretion at  $z\sim2.5$, when most orbital parameters differ from their $z=0$ values by $\sim20\%$.
This suggests that the above results are driven more strongly by evolving gas physics than by the nature of mass accretion.

Our results also have clear implications for satellite galaxy evolution within groups/clusters.
For example, using the parametrization of satellite survival time given in  \citet{BoyMaQua08}, changes in circularity across mass and redshift from the global $z=0$ value can lead to reductions of satellite lifetimes of $30\%$ or more.
Additionally, environmental effects such as ram-pressure stripping of gas and tidal stripping of stars are expected to occur primarily at orbital pericenter \citep[e.g.,][]{DekDevHet03,TayBab04,McCFreFon08}.
Beyond their dependence on evolving gas physics, our results suggest that satellite galaxy quenching and morphological evolution proceed more efficiently and rapidly at higher group/cluster masses and higher redshift.

While this work focuses on the orbital parameters of satellite halos at the time of infall, it is not immediately clear how well these orbital distributions and their mass and redshift dependencies persist to satellite populations well after infall, as the orbits become affected by dynamical friction, tidal stripping, and extended halo potentials (see the Appendix for some estimate of the latter).
There has been some work examining satellite subhalo orbits within host halos \citep{GilKneGib04,ReeGovQui05,SalNavAba07}, which found orbital parameter distributions similar to those here (for example, $\eta\approx0.5$).
This lack of circularity evolution after infall is plausible since dynamical friction is expected to be inefficient in altering orbital parameters such as circularity \citep{vdBLewLak99}.
Furthermore, \citet{Fal10} recently found that the orbits of satellite subhalos within host halos at $z=0$ are more radially biased in more massive host halos, which suggests that the results here remain valid well after infall.
We will pursue a more robust investigation into the evolution of satellite orbits after infall in future work.

There is also possible observational evidence in support of the trends seen here.
\citet{HerZarJin08} examined the orbital velocities of galaxies in local galaxy clusters, finding highly asymmetric velocity distributions consistent with satellites largely retaining their infalling orbits.
Promisingly, \citet{BivPog09} examined satellite galaxy orbits in galaxy clusters from $z=0$ to $z=0.8$ and found evidence that satellite orbits are indeed less isotropic (more radial) at higher redshift.

Finally, the results on satellite velocities also have implications for relating satellite dynamics to those of the overall group/cluster.
At lower host halo masses, satellite velocities become significantly `hotter' than than the host halo, implying a possible systematic biasing in using satellite velocity dispersions to infer halo masses \citep[][found a similar trend with mass for the velocity bias of satellite subhalos within host halos]{Fal10}.
Future work will involve a more detailed analysis of satellite velocity bias and its mass and redshift dependence.

\section*{Acknowledgments}

I gratefully acknowledge the support of an NSF Graduate Research Fellowship.
I thank Martin White for use of simulation data, Avi Loeb, Neal Dalal, and Frank van den Bosch for useful conversations, and Martin White and Joanne Cohn for comments on an early draft.
The simulation was run at the National Energy Research Scientific Computing Center.

\bibliography{ms}

\appendix
\section{Impact of Halo Profile on Orbits after Infall}

\begin{figure}
\begin{center}
\resizebox{3in}{!}{\includegraphics{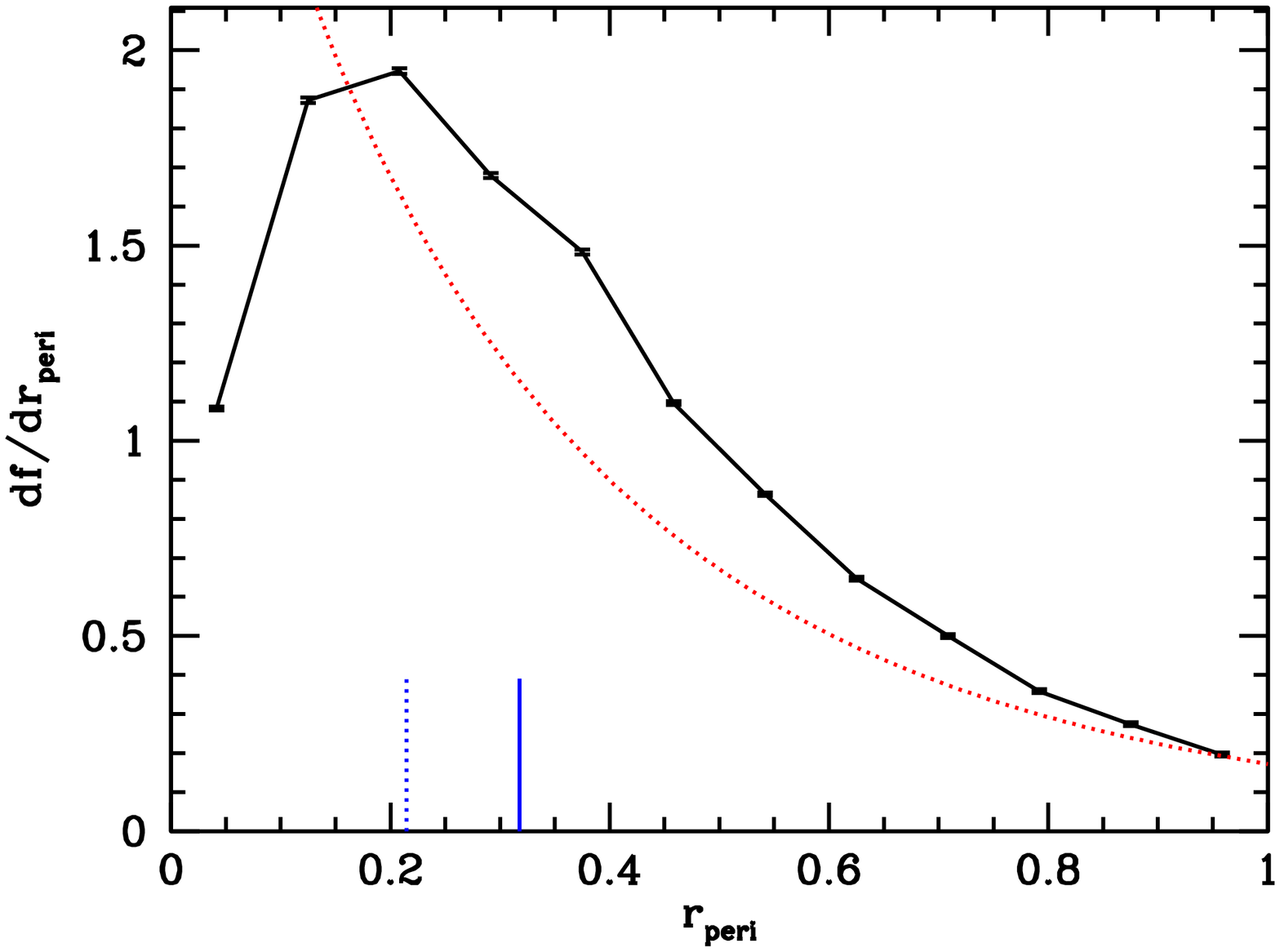}}
\end{center}
\vspace{-0.1in}
\caption{
Distribution of pericenter (scaled to host halo virial radius) for infalling satellites halos at $z=0$, for all halos $>10^{10}\hmsol$.
Solid curves shows distribution derived by numerically integrating satellite orbits after infall assuming a spherical NFW density profile for the host halo.
For comparison, dotted curve shows distribution derived via the point particle approximation, as in Fig.~\ref{fig:circ_rperi_dist}.
Vertical lines show corresponding median values.
} \label{fig:rperi_dist_nfw}
\end{figure}

Satellite orbital circularity and pericenter as explored in this work were derived from the point particle approximation and so are only strictly valid at the time of infall.
Circularity and pericenter are not generally conserved for orbits within more realistic extended halo potentials.
In particular, the actual pericentric distance that a satellite experiences after infall can differ systematically from the value derived at infall.
Here, we briefly explore the impact of an extended host halo profile on the satellite pericenter distribution at $z=0$.
To do so, we proceeded in the same manner as above (outlined in \S\ref{sec:methods}), except that we modeled host halos as having spherical NFW profiles as given by their measured concentrations, and we numerically integrated the orbits of (point particle) satellites to obtain their distance of closest approach.

Figure~\ref{fig:rperi_dist_nfw} shows this pericenter distribution at $z=0$, for all halos $>10^{10}\hmsol$.
For comparison, the dotted curve shows the original distribution that was obtained from the point particle approximation (same as in Fig.~\ref{fig:circ_rperi_dist}).
Because an extended halo profile contains less mass at smaller radii, a satellite experiences reduced gravitational focusing as it nears halo center, causing the pericenter distribution to roll over at small radii.
Overall, the spherical NFW pericenter distribution is skewed to higher values, causing the median to increase from $0.21$ (point particle) to $0.32$ (NFW).

Thus, the pericenter distributions given in this paper may moderately underestimate the true distances of closest approach that satellites experience.
However, beyond this systematic offset, we find that using an NFW profile produces no significant change in the mass and redshift trends seen above.
Furthermore, other factors such as triaxial profiles, dynamical friction, mass stripping, and scattering processes will also alter a satellite's orbital properties after infall, and future work will explore these in more detail.

\label{lastpage}

\end{document}